\documentclass[aps,prl,amsmath,amssymb,twocolumn,superscriptaddress,showpacs]{revtex4-1}
\usepackage{amsmath,amssymb}
\usepackage{graphicx}
\usepackage{appendix}
\usepackage[usenames,dvipsnames]{color}
\usepackage[colorlinks=true, citecolor=blue, linkcolor=blue, urlcolor=blue]{hyperref}

\usepackage[nottoc,numbib]{tocbibind}

\newcommand{\vk}{{\vec{k}}}
\newcommand{\va}{{\hat{\vec{A}}}}
\newcommand{\ve}{{\vec{e}}}

\begin{document}

\title{Cavity Quantum-Electrodynamical Chern Insulator: \\
Route Towards Light-Induced Quantized Anomalous Hall Effect in Graphene}

\author{Xiao Wang}
\affiliation 
{Department of Physics, Tsinghua University, Beijing 100084, China}

\author{Enrico Ronca}
\affiliation 
{Max Planck Institute for the Structure and Dynamics of Matter, Luruper Chaussee 149, 22761 Hamburg, Germany}

\author{Michael A.~Sentef}
\email{michael.sentef@mpsd.mpg.de}
\affiliation 
{Max Planck Institute for the Structure and Dynamics of Matter, Luruper Chaussee 149, 22761 Hamburg, Germany}

\date{\today}

\begin{abstract}
We show that an energy gap is induced in graphene by light-matter coupling to a circularly polarized photon mode in a cavity. Using many-body perturbation theory we compute the electronic spectra which exhibit photon-dressed sidebands akin to Floquet sidebands for laser-driven materials. In contrast with Floquet topological insulators, in which a strictly quantized Hall response is induced by light only for off-resonant driving in the high-frequency limit, the photon-dressed Dirac fermions in the cavity show a quantized Hall response characterized by an integer Chern number. Specifically for graphene we predict that a Hall conductance of $2 e^2/h$ can be induced in the low-temperature limit.
\end{abstract}
\maketitle

The tunability of the properties of matter by light-matter coupling is becoming a unifying scheme across many disciplines, ranging from pump-probe spectroscopies \cite{giannetti_ultrafast_2016,basov_towards_2017,kemper_review_2017,cavalleri_photo-induced_2018} via artificial gauge fields in cold atoms \cite{struck_quantum_2011,aidelsburger_experimental_2011,miyake_realizing_2013,jotzu_experimental_2014} to strong light-matter coupling in polaritonic chemistry \cite{ruggenthaler_quantum-electrodynamical_2018,vaidya_tunable-range_2018,ribeiro_polariton_2018,flick_strong_2018,kockum_ultrastrong_2019}. Floquet- and nonequilibrium topological states of matter have been of particular theoretical interest \cite{oka_photovoltaic_2009,lindner_floquet_2011,kitagawa_transport_2011,rechtsman_photonic_2013,usaj_irradiated_2014,dehghani_dissipative_2014,claassen_all-optical_2016,hubener_creating_2017,schuler_quench_2018} but reports of theoretically predicted Floquet-band formation in time-resolved photoemission \cite{sentef_theory_2015} in solids are still rare \cite{wang_observation_2013,mahmood_selective_2016}. Only recently an anomalous Hall effect induced by circularly polarized light has been reported in graphene \cite{mciver_light-induced_2018}. One of the major hurdles towards controlling interesting phases of matter with classical light in solids lies in heating effects that typically hide the low-energy properties of Floquet-engineered Hamiltonians and prevent quantized topological properties, unless very specific setups are considered \cite{foa_torres_multiterminal_2014,rudner_anomalous_2013}. Therefore the manipulation of many-body systems with quantum instead of classical light is a topic of increasing interest \cite{laussy_exciton-polariton_2010,cotlet_superconductivity_2016,kavokin_excitonpolariton_2016,hagenmuller_cavity-enhanced_2017,mivehvar_superradiant_2017,sentef_cavity_2018,rosner_plasmonic_2018,schlawin_cavity-mediated_2018,mazza_superradiant_2019,curtis_cavity_2018,kiffner_manipulating_2019,allocca_cavity_2019,rokaj_quantum_2018,latini_cavity_2018}. 
In quantum optics the potential of chiral light-matter coupling has been recognized to bear potential for instance for the design of ultrafast optical switches or nonreciprocal devices \cite{lodahl_chiral_2017}, and chiral exciton-plasmon coupling has been demonstrated experimentally \cite{chervy_room_2018}. Recently chiral light-matter coupling in cavities has been suggested to reverse the sign of the Casimir force between the metallic plates \cite{jiang_chiral_2019}.

\begin{figure}[ht!] 
  \begin{center}
    \includegraphics[trim={0 6cm 0 6cm},clip,width=\columnwidth]{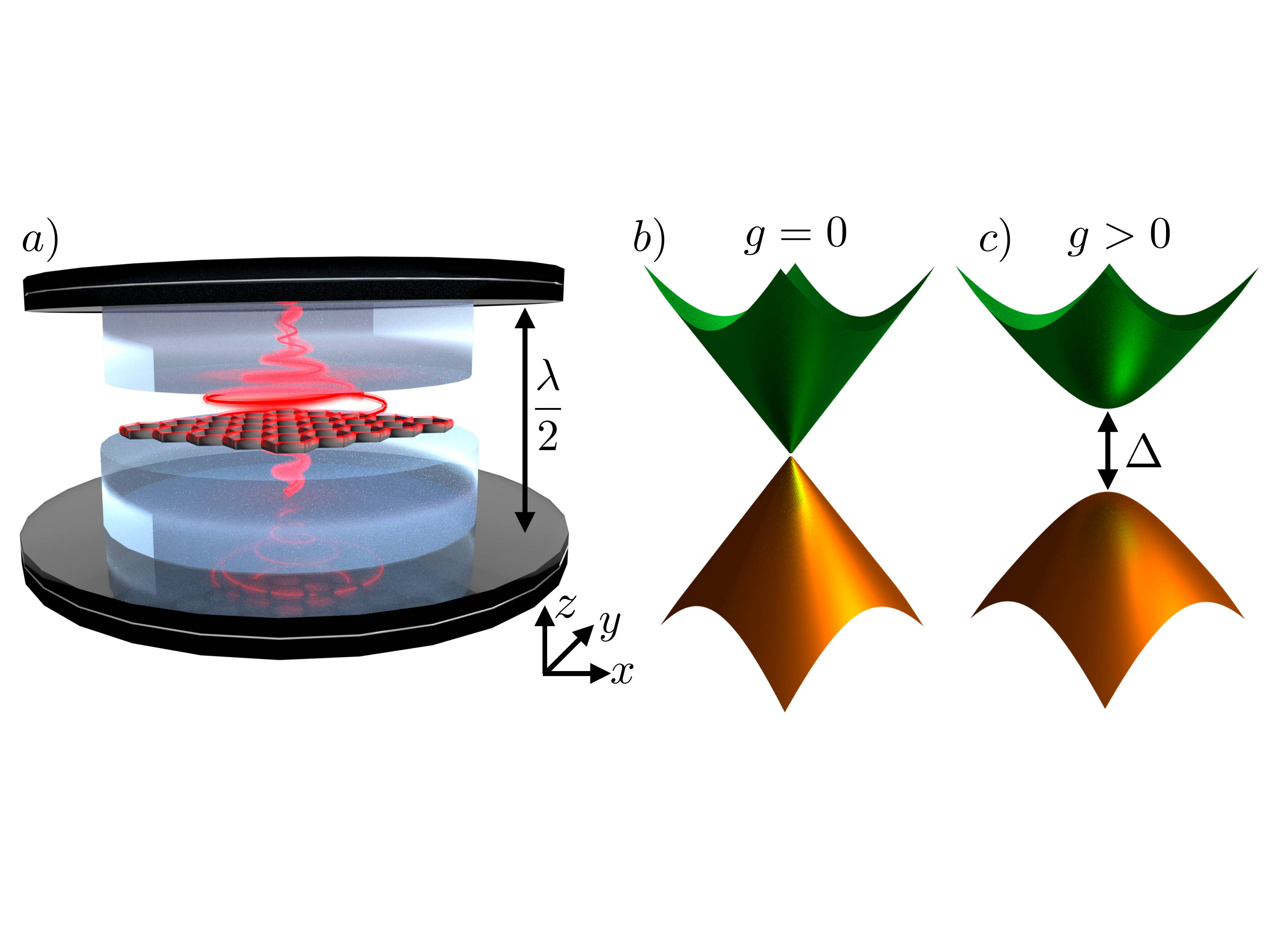}
    \caption{
    \label{fig:cavity} 
    {\bf Two-dimensional (2D) material inside a chiral cavity.} (a) Setup for 2D graphene between cavity mirrors a distance $\lambda/2$ apart, where $\lambda$ is the wavelength of the fundamental cavity photon mode. The red spiral indicates the circular photon polarization. The 2D material is encapsulated in a dielectric medium (glassy region). (b) Dirac cone of a 2D Dirac material at electron-photon coupling $g=0$. (c) Energy gap $\Delta$ due to time-reversal symmetry breaking for $g>0$. 
    }
  \end{center}
\end{figure}

Here we show that coupling two-dimensional Dirac fermions to circularly polarized light in a quantum-electrodynamical (QED) cavity (Fig.~\ref{fig:cavity}(a)) gives rise to an energy gap $\Delta$ at the Dirac point (Fig.~\ref{fig:cavity}(b),(c)). This gap opening is due to the breaking of time-reversal symmetry in close analogy to the Floquet case of circularly polarized classical light \cite{oka_photovoltaic_2009} and closely inspired by Haldane's original proposal of chiral hopping on a honeycomb lattice \cite{haldane_model_1988}. Within many-body perturbation theory we show that the scaling of the light-induced energy gap with light-matter coupling strength and light frequency is analogous to the high-frequency limit in the classical case. Importantly, in contrast to the classical case, this happens in the ground state of the cavity, and therefore without electronic excitations across the energy gap when the experiment is performed at sufficiently low temperature. On top of that the light-matter coupling also induces photon-shakeoff sidebands in the electronic spectrum, again in close analogy to the Floquet case. While a band gap induced by chiral vacuum fluctuations was suggested earlier \cite{kibis_band_2011}, it was not connected to light-induced topology. Here we predict that for graphene the chiral light-matter coupling can give rise to a quantized anomalous Hall effect (QAHE) \cite{chang_experimental_2013} at accessible temperatures.

We consider a two-sublattice ($A$, $B$) Hamiltonian with inter-sublattice hybridizations $\gamma(\vk)$, to be specified below, minimally coupled to a single QED cavity photon mode,
\begin{align}
H =& \sum_\vk  \left(
\begin{array}{c}
c^{\dagger}_{A,\vk}\\
c^{\dagger}_{B,\vk}
\end{array} 
\right)^T
\left(
\begin{array}{cc}
0 & \gamma(\vk-\va) \\
\gamma(\vk-\va)^{\dagger} & 0 
\end{array}
\right)
\left(
\begin{array}{c}
c^{}_{A,\vk}\\
c^{}_{B,\vk}
\end{array} 
\right) \nonumber \\ + &\sum_\lambda \omega_\lambda a^{\dagger}_\lambda a^{}_\lambda,
\label{eq:ham}
\end{align}
where we use the dipole approximation and assume coupling only at zero momentum transfer to photons with vector potential
\begin{align}
\va &= A_0 \sum_\lambda \left(\ve_\lambda a^{}_\lambda + \ve_\lambda^* a^{\dagger}_\lambda\right),
\end{align}
where $A_0 = \sqrt{\hbar/(\epsilon \epsilon_0 V \omega)}$, with vacuum permittivity $\epsilon_0$, cavity volume $V$, and dielectric constant $\epsilon$ of the dielectric embedding of a two-dimensional material inside the cavity. Here we have assumed an infinitely extended cavity in the $x$-$y$ plane for simplicity.
For a single two-dimensional spinless Dirac fermion with Fermi velocity $v_F$ we have
$
\gamma(\vk) = \hbar v_F (k_x + i k_y).
$
Notice that the $A^2$ term that usually appears for massive charged particles minimally coupled to a gauge field is absent for the massless Dirac fermions considered here. 

Using a right-handed circularly polarized cavity reduces the photon field to a single branch with $\ve_\lambda \equiv \ve$, operators $a^{\dagger}_\lambda \equiv a^{\dagger}$, and frequency $\omega_\lambda \equiv \omega$, with unit polarization vector
$
\ve = \frac{1}{\sqrt2} (1,i).
$
In this case $\gamma(\vk-\va) \rightarrow \hbar v_F (k_x + i k_y - \sqrt2 A_0 a^{\dagger})$ in Eq.~(\ref{eq:ham}).  

In the following we investigate the photon dressing effects on the electronic structure by means of many-body perturbation theory using Matsubara Green's functions. To lowest order in the effective electron-photon coupling strength $g \equiv v_F A_0 \sqrt2$ we obtain the zero-temperature, energy-dependent retarded electronic self-energy at the Dirac point $\vk = 0$ within the non-selfconsistent first Born approximation \cite{wang_supplementary_nodate} as 
\begin{align}
\Sigma^R_{0,aa}(\vk=0, \epsilon) &= \frac{g^2/2}{\epsilon + i 0^+ -\omega}, \label{eq:selfaa} \\
\Sigma^R_{0,bb}(\vk=0, \epsilon) &= \frac{g^2/2}{\epsilon + i 0^+ +\omega},  \label{eq:selfbb} 
\end{align}
where $aa$ and $bb$ refer to the intra-sublattice self-energies on sublattices $A$ and $B$, respectively.  

The relevant quantity to analyze renormalizations of the electronic structure due to light-matter coupling is the electronic single-particle spectral function
\begin{align}
A(k, \epsilon) &= -\frac{1}{\pi} \text{Im} \text{Tr} \hat{G}^R(k, \epsilon),  \label{eq:intensity}
\end{align}
obtained from the retarded Green's function on the real-energy axis, which is related to the self-energy via the Dyson equation
$
\hat{G}^{R,-1}(k, \epsilon) = \hat{G}_0^{R,-1}(k, \epsilon) - \hat{\Sigma}^R (k, \epsilon)
$. In the following we discuss both the non-selfconsistent, lowest order self-energies $\Sigma_0 \propto g^2 G_0 D_0$ with bare electron ($G_0$) and photon ($D_0$) propagators, as well as electronically self-consistent $\Sigma \propto g^2 G D_0$ (see \cite{wang_supplementary_nodate} for details) with dressed electronic Green's function $G$ obtained from the Dyson equation. The equations for $\Sigma[G]$ and $G[\Sigma]$ are solved self-consistently until convergence is reached. We neglect dressing of the photon propagators within this work, as its main effect is a slight renormalization of the photon frequency and an acquisition of a finite photon lifetime. In particular, the photon renormalization leads to a correction of order $g^4$ to the electronic self-energy. As we are interested here in the realistic scenario of weak light-matter coupling, we expect that these effects will be small and will not qualitatively affect our results and conclusions.

Considering the lowest order self-energy at the Dirac point $\hat{\Sigma}^R (k=0, \epsilon) \rightarrow \hat{\Sigma}^R_0 (k=0, \epsilon)$ from Eqs.~(\ref{eq:selfaa},\ref{eq:selfbb}) we find that $A(k=0, \epsilon)$ acquires an energy gap $\Delta = \sqrt{2g^2+\omega^2}-\omega$. In the limit $2g^2/\omega^2 \ll 1$ we obtain
$\Delta \approx \frac{g^2}{\omega} = \frac{2 \hbar^2 v_F^2 A_0^2 }{\omega}$. This result is in remarkably close formal analogy with the Floquet high-frequency expansion \cite{oka_photovoltaic_2009} when the quantum photon amplitude $A_0$ is replaced by the field strength $A_0$ of the classical vector potential. 

\begin{figure}[htp!] 
  \begin{center}
    \includegraphics[width=1.0\columnwidth]{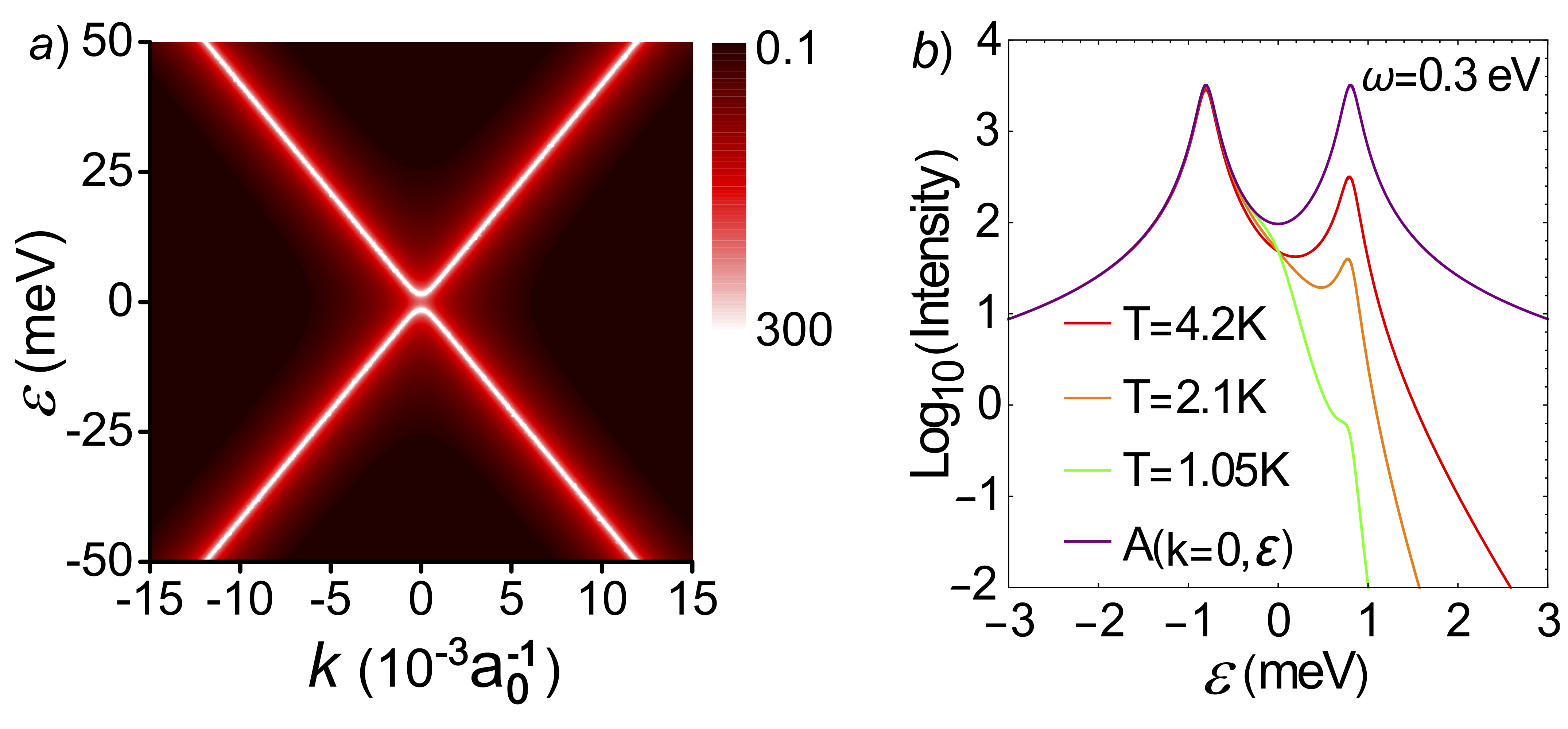}
    \caption{
    \label{fig:arpes} 
    {\bf Electronic spectral function in circularly polarized cavity.}  (a) Spectral intensity $A(k, \epsilon)$ versus momentum $k$ and binding energy $\epsilon$ for temperature $T = 4.2$ K and cavity frequency $\omega = 0.3$ eV for coupling strength $g=0.023$ eV. (b) Line cut of the spectral intensity $A(k, \epsilon)$ at the Dirac point $k=0$ for the same parameters as in (a) on a logarithmic scale. Colored lines show the occupied part of the spectrum at different temperatures.
    }
  \end{center}
\end{figure}

In order to estimate the coupling strength in a realistic device, we consider graphene encapsulated in hexagonal boron nitride with dielectric constant $\epsilon \approx 7$ for in-plane light polarization inside a plasmonic cavity. We obtain $g[\text{eV}]=(\hbar v_F)[\text{eV} a_0] \sqrt{4 \alpha \lambda /(\epsilon V)}$ and use the effective cavity volume $V = 2.5 \times 10^{-5} \times (\lambda/(2\sqrt{\epsilon}))^3$ \cite{maissen_ultrastrong_2014,schlawin_cavity-mediated_2018} with the photon wavelength $\lambda$ being twice the cavity size in $z$ direction. Lengths are measured in units of the graphene interatomic distance $a_0=1.42 \text{\AA}$, $\hbar v_F = 4.2 \text{eV} a_0$, and $\alpha \approx 1/137$ is the fine structure constant. With these values we obtain $g \approx 0.0077 \text{eV}$ for $\omega = 0.1 \text{eV}$ (cavity size $6.2 \mu\text{m}$) and $g \approx 0.023 \text{eV}$ for $\omega = 0.3 \text{eV}$ ($2.07 \mu\text{m}$). Since these values are still safely in the weak-coupling regime the corresponding energy gaps are given by $g^2/\omega$ and take values of 0.00059 eV and 0.0018 eV, respectively. 

\begin{figure*}[ht!]
\centering
\includegraphics[width=0.8\textwidth]{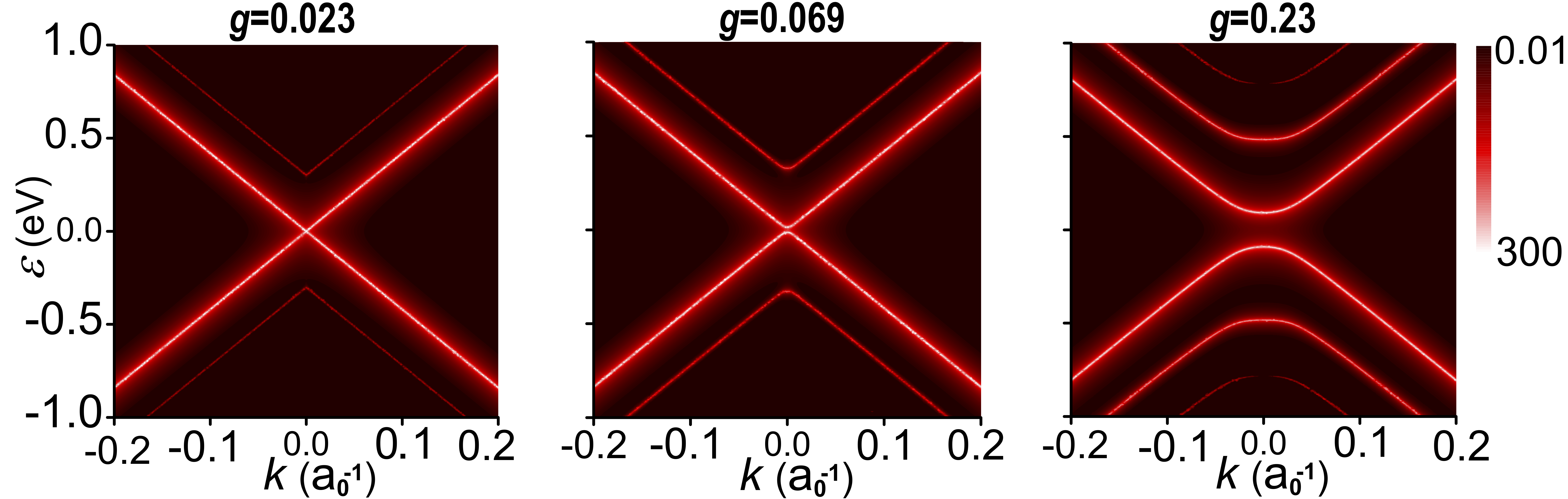}
    \caption{
    \label{fig:sidebands} 
    {\bf Spectral function showing sidebands for varying coupling strength.} (a) Spectral intensity $A(k, \epsilon)$ (Dirac point) for temperature $T = 1.0$ K and cavity frequency $\omega = 0.3$ eV at coupling strength $g=0.069$ eV. (b,c) Same as in (a) for increased coupling strengths $g=0.069$ eV and $g=0.23$ eV, respectively.
    }
\end{figure*}

We go beyond the lowest-order perturbation theory by numerically solving for the selfconsistent self-energy $\Sigma$ on the Matsubara frequency axis and obtain the real-frequency self-energy by Pad\'e approximants \cite{beach_reliable_2000}. We checked our procedure against the closed-form analytical continuation result for the non-selfconsistent case. The resulting electronic spectral function $A(k, \epsilon)$ is shown in Fig.~\ref{fig:arpes}(a) for a cavity with photon frequency $\omega = 0.3$ eV. Indeed a small energy gap is obtained, as discussed above. In order to better resolve this gap at the Dirac point, we show in Fig.~\ref{fig:arpes}(b) a line cut of the same data as in Fig.~\ref{fig:arpes}(a). 

Importantly, the filling of electronic states within many-body perturbation theory in thermal equilibrium is simply given by a Fermi-Dirac distribution at the temperature of interest. To show the thermal occupation effect, we plot in Fig.~\ref{fig:arpes}(b) 
the occupied electronic spectra $f(\epsilon, T) A(k, \epsilon)$ with Fermi function $f(\epsilon, T) = 1/(\exp(\epsilon/(k_B T))+1)$
for cryostatic temperatures of 4.2 K, 2.1 K, and 1.05 K. 
Indeed at sufficiently low temperature one obtains basically filled states in the valence band below zero energy and empty states in the conduction band above. This will be important for the discussion of the QAHE below.

As a next step we are interested in the larger-scale renormalization of the electronic structure by photon dressing. In the classical laser-driving case one obtains Floquet sidebands due to emission and absorption of photons from the laser field, and these sidebands can be measured in time- and angle-resolved photoemission spectroscopy \cite{wang_observation_2013, sentef_theory_2015}.
Fig.~\ref{fig:sidebands} shows the electronic spectra $A(k, \epsilon)$ for $\omega=0.3$ eV at varying coupling strength: $g=0.023$ (same as in Fig.~\ref{fig:arpes}), $g=0.069$, and $g=0.23$. At $g=0.023$ only the first photon sideband is barely visible here, separated from the main band by the photon frequency. As the coupling strength is increased to $g=0.069$ the sideband becomes more pronounced. On the energy scale required to see sidebands the energy gap at the Dirac point is hardly visible. For the strongest coupling $g=0.23$ the second-order sideband becomes also visible, and the energy gap is pronounced. Notice that these sidebands look quite different from the classically driven Floquet case (cf.~Ref.~\onlinecite{sentef_theory_2015}) because in the undriven cavity the rates of photon absorption and emission are obviously very different, especially at weak coupling and low temperature when the photons are almost in their vacuum state. By contrast, in the driven case the photons are basically in a coherent state with large occupation numbers.

We now turn to the discussion of the light-induced anomalous Hall effect.
As outlined above the occupied electronic spectrum is given by the equilibrium Fermi-Dirac distribution, which is in marked contrast from the driven Floquet case. In particular, at zero temperature only electronic states below the Fermi energy, $\epsilon=0$, are filled and those above are empty. This immediately implies that the Hall conductance can be computed from the Chern number of the dressed electronic structure. This can be achieved by employing the topological invariant computed from the interacting Green's function matrix of a 2+1-dimensional system \cite{ishikawa_magnetic_1986,volovik_universe_2003,wang_topological_2010}
\begin{align}
N_2 &=
\frac{1}{24 \pi^2}
\int dk_0 dk_y dk_y 
\text{Tr}\left[\epsilon^{\mu\nu\rho}\hat{G}\partial_\mu \hat{G}^{-1}\hat{G} \partial_\nu \hat{G}^{-1} \hat{G} \partial_\rho \hat{G}^{-1} \right],
\end{align}
where $\mu$, $\nu$, $\rho$ run through $k_0$, $k_x$, $k_y$ and $k_0 = i \omega$ is imaginary frequency. The invariant $N_2$ is equal to the first Chern number $C_1$, of an effectively noninteracting system with Hamiltonian $\hat{h}_\text{eff}(\vk) = -\hat{G}^{-1}(\vk,0)$. 

\begin{figure}[ht!] 
  \begin{center}
    \includegraphics[width=\columnwidth]{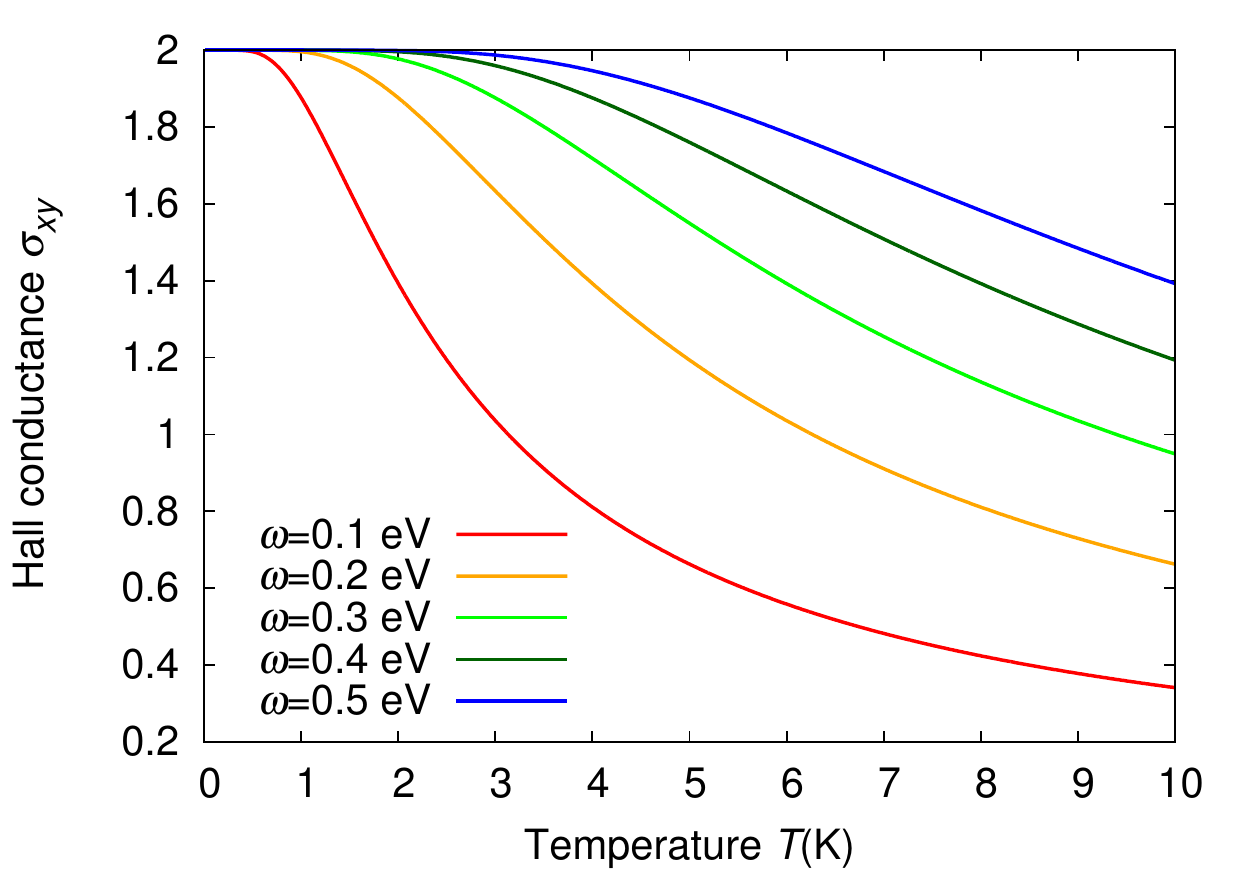}
    \caption{
    \label{fig:hall} 
    {\bf Hall conductance for different cavity frequencies.} Hall conductance as a function of temperature for cavity photon frequencies as indicated.
    }
  \end{center}
\end{figure}

For the single photon-dressed Dirac fermion we obtain $N_2 = \pm \frac12$ for the choice of right-handed ($+$) or left-handed ($-$) photon polarization. In graphene, one has two Dirac fermions in the Brillouin zone of opposite chirality, with $\gamma(\vk) = v_F (k_x \pm i k_y)$ and the sign $\pm$ referring to the Dirac-fermion chirality. Repeating the calculations for the negative-chirality Dirac fermion, one obtains the same contribution to the topological invariant for given photon chirality. Therefore, we obtain the final result for the zero-temperature Hall conductance of spinfull graphene in a circularly polarized cavity
\begin{align}
\sigma_{xy} &= \pm 2 \frac{e^2}{h}
\end{align}
with the sign $\pm$ determined only by the photon chirality, and $2 = \frac12 \times N_s \times N_v$ with spin degeneracy $N_s=2$ and valley degeneracy $N_v=2$. Fig.~\ref{fig:hall} shows the Hall conductance with a right-handed circularly polarized photon mode obtained for different cavity frequencies and their respective coupling strengths as a function of temperature by multiplying the respective Berry curvatures with the Fermi-Dirac distribution. For each of the frequencies, there is a characteristic gap energy scale below which the Hall conductance quickly approaches the quantized limit. Importantly the fully quantized limit is reached at cryostatic temperatures that can be reached in the laboratory.

Finally, we briefly discuss similarities and differences with respect to the Floquet case of classical circularly polarized light. The most important similarities are the breaking of time-reversal symmetry leading to an energy gap at the Dirac point and the scaling of the gap with field amplitude and frequency. The key difference is that the gap opening happens for a vanishing macroscopic electromagnetic field in the cavity, i.e., for a zero classical field. The effect here is purely based on photonic quantum fluctuations and happens at thermal equilibrium. For a classical field, unless the high-frequency limit or specific setups are considered \cite{foa_torres_multiterminal_2014,rudner_anomalous_2013}, heating effects usually destroy the quantized anomalous Hall conductance.

In summary, we have shown that a circularly polarized cavity gives rise to a quantized Hall conductance in two-dimensional Dirac fermions at zero temperature. Importantly, in practice the size of the energy gap will set the temperature scale below which the QAHE can be observed experimentally. Realistic estimates for plasmonic micro- and nanocavities yield temperatures of tens of Kelvin for the gap sizes, which implies that a QAHE induced by QED environments should be within experimental reach. We notice that electron-electron interactions in graphene should not destroy the proposed effect, as for weak coupling their main effect is a band renormalization. Regarding the potential detrimental role of disorder, we remark that graphene samples encapsulated in boron nitride, as considered by us, show ultrahigh mobilities and a very low density of carrier inhomogeneities \cite{dean_boron_2010}, which should allow for the observation of the topological gap at sufficiently low temperatures.

Similar ideas to push from the classical Floquet regime to the quantized collectively coupled Dicke regime have been put forward \cite{gulacsi_floquet_2015}. Natural follow-up questions pertain to the crossover from the classical to the QED limit, which could find interesting applications for instance to light-controlled topological superconductivity \cite{thakurathi_floquet_2017,dehghani_dynamical_2017,claassen_universal_2018}. Similarly it will be interesting to investigate how effective couplings in materials, such as dynamical Hubbard $U$ \cite{singla_thz-frequency_2015,tancogne-dejean_ultrafast_2018,topp_all-optical_2018,golez_dynamics_2018}, the exchange interaction \cite{mentink_ultrafast_2015,coulthard_enhancement_2017,kiffner_manipulating_2019}, or electron-phonon couplings \cite{knap_dynamical_2016,kennes_transient_2017,sentef_light-enhanced_2017,murakami_nonequilibrium_2017}, can be affected by quantum rather than classical light.

\textit{Acknowledgment.} 
Discussions with U.~de Giovannini, H.~H\"ubener, S.~Latini, A.~Rubio, and S.~A.~Sato are gratefully acknowledged.
X.W.~acknowledges support by the Tsinghua Xuetang Talents program. 
M.A.S.~acknowledges financial support by the DFG through the Emmy Noether program (SE 2558/2-1).

\bibliography{Cavity_Chern}

\onecolumngrid
\appendix

\setcounter{figure}{0}
\setcounter{section}{0}
\setcounter{equation}{0}
\makeatletter 
\renewcommand{\theequation}{S\@arabic\c@equation}
\makeatother
\makeatletter 
\renewcommand{\thefigure}{S\@arabic\c@figure}
\makeatother
\makeatletter 
\renewcommand{\thesection}{S\@arabic\c@section}
\makeatother

\section{Supplementary Material}
\subsection{Analytical calculation of lowest-order self-energy}
We derive the lowest-order perturbative correction to the single-particle electronic Hamiltonian via the Matsubara Green's function
\begin{align}
\hat{G}^{-1} = \hat{G}_0^{-1} - \hat{\Sigma}_0,
\end{align}
with the electronic Green's function in matrix form
\begin{align}
\hat{G}(\vk, \tau) &= -\mathcal{T}_\tau
\left(
\begin{array}{cc}
\langle c^{}_{A,\vk}(\tau) c^{\dagger}_{A,\vk}  \rangle & \langle c^{}_{A,\vk}(\tau) c^{\dagger}_{B,\vk}  \rangle \\
\langle c^{}_{B,\vk}(\tau) c^{\dagger}_{A,\vk}  \rangle & \langle c^{}_{B,\vk}(\tau) c^{\dagger}_{B,\vk}  \rangle 
\end{array}
\right).
\end{align}
The bare electronic Green's function written in Matsubara frequency space is
\begin{align}
\hat{G}_0(\vk, i p_n) &= (i p_n \hat{1} - \hat{h}_0(\vk))^{-1} =
\left(
\begin{array}{cc}
i p_n & -v_F(k_x + i k_y) \\
-v_F(k_x - i k_y) & i p_n  
\end{array}
\right)^{-1}.
\end{align}
In particular we have for the diagonal elements in orbital-momentum basis
\begin{align}
G_{0,aa}(\vk, i p_n) = G_{0,bb}(\vk, i p_n) = \frac{-i p_n}{p_n^2 + v_F^2 k^2}.
\end{align}
The lowest order perturbative self-energy is then given by 
\begin{align}
\Sigma_{0,aa}(\vk, i p_n) &= -\frac{g^2}{\beta} \sum_m \frac{-1}{i \omega_m + \omega} G_{0,bb}(\vk, i p_n + i \omega_m), \label{eq:saa} \\
\Sigma_{0,bb}(\vk, i p_n) &= -\frac{g^2}{\beta} \sum_m \frac{1}{i \omega_m - \omega} G_{0,aa}(\vk, i p_n + i \omega_m), \label{eq:sbb}
\end{align}
or explicitly
\begin{align}
\Sigma_{0,aa}(\vk, i p_n) &= -\frac{g^2}{\beta} \sum_m \frac{-1}{i \omega_m + \omega} \frac{(-i p_n - i \omega_m)}{(p_n + \omega_m)^2 + v_F^2 k^2}, \label{eq:sigmaaa}\\
\Sigma_{0,bb}(\vk, i p_n) &= -\frac{g^2}{\beta} \sum_m \frac{1}{i \omega_m - \omega} \frac{(-i p_n - i \omega_m)}{(p_n + \omega_m)^2 + v_F^2 k^2}, \label{eq:sigmabb}
\end{align}
with $g \equiv v_F A_0 \sqrt2$.
Here we have used the photon Green's functions
$D_1(\tau) = -\mathcal{T}_\tau \langle a^{\dagger}(\tau) a^{}  \rangle$ and $D_2(\tau) = -\mathcal{T}_\tau \langle a^{}(\tau) a^{\dagger} \rangle$, for which the bare propagators in Matsubara frequency space read $D_{1,0}(i \omega_m) = \frac{-1}{i \omega_m + \omega}$ and $D_{2,0}(i \omega_m) = \frac{1}{i \omega_m - \omega}$, respectively.

The only remaining task is to carry out the Matsubara summation as follows. Consider the summation for $\Sigma_{0,aa} = -g^2 S$, with
\begin{align}
S &= \frac{1}{\beta} \sum_m \frac{1}{i \omega_m + \omega} \frac{(i p_n + i \omega_m)}{(p_n + \omega_m)^2 + v_F^2 k^2}.
\end{align}
This is written as 
\begin{align}
S &= -\frac{1}{\beta} \sum_m f(i \omega_m),
\end{align}
which is evaluated by a contour integration
\begin{align}
I &= \lim_{R \rightarrow \infty} \oint \frac{dz}{2 \pi i} f(z) n_B(z),
\end{align}
with the Bose function $n_B(z) \equiv \frac{1}{e^{\beta z}-1}$, and
\begin{align}
f(z) &\equiv \frac{1}{z + \omega} \frac{(i p_n + z)}{(i p_n + z)^2 - v_F^2 k^2}.
\end{align}
We use the residual theorem for first order poles of $g(z) \equiv f(z) n_B(z)$ with residues given by
\begin{align}
R &= \lim_{z \rightarrow z_0} (z-z_0) g(z).
\end{align}
The poles of the integrand of $I$ and their respective residues are
\begin{align}
\tilde{z}_m &= i 2 \pi m/\beta, \; R_m = \frac{1}{\beta}f(i \omega_m), \\
z_1 &= - \omega, \; R_1 = \frac{(i p_n - \omega)}{(i p_n -\omega)^2- v_F^2 k^2} n_B(-\omega),\\
z_{2,3} &= -i p_n \pm v_F k, \; R_{2,3} = \frac{1}{\pm z_2 + \omega} \lim_{z \rightarrow \pm z_2} \frac{(i p_n + z)(z - z_2)}{(i p_n + z)^2 - v_F^2 k^2} n_B(z). 
\end{align}
A straightforward calculation gives
\begin{align}
R_{2/3} &= \frac{1}{-i p_n \pm v_F k + \omega} \frac{n_B(-i p_n \pm v_F k)}{2}.
\end{align}
Using $e^{-\beta i p_n} = -1$ for all fermionic Matsubaras $p_n$ these residues can be written as
\begin{align}
R_{2/3} &= \frac{1}{i p_n \mp v_F k - \omega} \frac{n_F(\pm v_F k)}{2},
\end{align}
and their sum
\begin{align}
R_2 + R_3 &= \frac12 \frac{(i p_n - \omega + v_F k)n_F(v_F k) + (i p_n - \omega - v_F k)(1-n_F(v_F k))}{(i p_n -\omega)^2- v_F^2 k^2}.
\end{align}
The total integral is then given by
\begin{align}
I &= \frac{1}{\beta} \sum_m f(i \omega_m) + \frac{(i p_n - \omega)n_B(-\omega) + \frac12 \left[(i p_n - \omega + v_F k)n_F(v_F k) + (i p_n - \omega - v_F k)(1-n_F(v_F k))\right]}{(i p_n -\omega)^2- v_F^2 k^2}.
\end{align}
The integral vanishes for $R \rightarrow \infty$, $I=0$, which gives the result
\begin{align}
S &= \frac{(i p_n - \omega)n_B(-\omega) + \frac12 \left[(i p_n - \omega + v_F k)n_F(v_F k) + (i p_n - \omega - v_F k)(1-n_F(v_F k))\right]}{(i p_n -\omega)^2- v_F^2 k^2}.
\end{align}
In summary, we have
\begin{align}
\Sigma_{0,aa}(\vk, i p_n) &= -g^2 \frac{(i p_n - \omega)n_B(-\omega) + \frac12 \left[(i p_n - \omega + v_F k)n_F(v_F k) + (i p_n - \omega - v_F k)(1-n_F(v_F k))\right]}{(i p_n -\omega)^2- v_F^2 k^2}, \\
\Sigma_{0,bb}(\vk, i p_n) &= g^2 \frac{(i p_n + \omega)n_B(\omega) + \frac12 \left[(i p_n + \omega + v_F k)n_F(v_F k) + (i p_n + \omega - v_F k)(1-n_F(v_F k))\right]}{(i p_n +\omega)^2- v_F^2 k^2},
\end{align}
where we have used that $\Sigma_{0,bb}$ is the negative of $\Sigma_{0,aa}$ with $\omega \rightarrow -\omega$ (cf.~Eqs.~(\ref{eq:sigmaaa}) and (\ref{eq:sigmabb})).

We analytically continue the Matsubara result to the real axis, $i p_n \rightarrow \epsilon + i 0^+$ to obtain the retarded self-energy
\begin{align}
\Sigma^R_{0,aa}(\vk, \epsilon) &= -g^2 \frac{(\epsilon + i 0^+ - \omega)n_B(-\omega) + \frac12 \left[(\epsilon + i 0^+ - \omega + v_F k)n_F(v_F k) + (\epsilon + i 0^+ - \omega - v_F k)(1-n_F(v_F k))\right]}{(\epsilon + i 0^+ -\omega)^2- v_F^2 k^2}, \\
\Sigma^R_{0,bb}(\vk, \epsilon) &= g^2 \frac{(\epsilon + i 0^+ + \omega)n_B(\omega) + \frac12 \left[(\epsilon + i 0^+ + \omega + v_F k)n_F(v_F k) + (\epsilon + i 0^+ + \omega - v_F k)(1-n_F(v_F k))\right]}{(\epsilon + i 0^+ +\omega)^2- v_F^2 k^2}.
\end{align}

At zero temperature $\beta \rightarrow \infty$ one has for $\omega > 0$ that $n_B(\omega) = 0$ and $n_B(-\omega) = -1$. Moreover at the Dirac point $k=0$ one has $n_F(0) = \frac12$ independent of temperature. Therefore at zero temperature and $k=0$
\begin{align}
\Sigma^R_{0,aa}(0, \epsilon) &= -g^2 \frac{(\epsilon + i 0^+ - \omega)(-1) + \frac12 (\epsilon + i 0^+ - \omega)}{(\epsilon + i 0^+ -\omega)^2} = \frac{g^2/2}{\epsilon + i 0^+ -\omega}, \\
\Sigma^R_{0,bb}(0, \epsilon) &= g^2 \frac{\frac12(\epsilon + i 0^+ + \omega)}{(\epsilon + i 0^+ +\omega)^2} = \frac{g^2/2}{\epsilon + i 0^+ +\omega}.
\end{align}

\subsection{Selfconsistent numerics}
Within electronically selfconsistent many-body perturbation theory, we evaluate at the lowest nonvanishing order ($g^2$) the Hartree and Fock self-energy diagrams on the Matsubara frequency axis. One can show that the Hartree diagrams vanish when using the bare photon propagator; in any case, the Hartree diagrams would only yield a static, non-retarded contribution to the self-energy. The nonvanishing Fock diagrams are the same as the non-selfconsistent ones discussed in detail above (Eqs.~(\ref{eq:saa},\ref{eq:sbb})), where we simply replace the bare electronic Green's function by the dressed ones $\hat{\Sigma}_0 \equiv \hat{\Sigma}[\hat{G}_0] \rightarrow \hat{\Sigma}[\hat{G}]$, which are obtained from the solution of the Dyson equation $\hat{G}^{-1} = \hat{G}_0^{-1} - \hat{\Sigma}$. 

\end{document}